\documentclass[aps,reprint,nofootinbib,prl]{revtex4-1}
\usepackage{graphicx}
\usepackage{amssymb}
\usepackage{epstopdf}
\usepackage{graphicx}
\usepackage{bm}
\usepackage{color}
\usepackage[caption=false]{subfig}
\usepackage{mhchem}

\newcommand{\be}{\begin{eqnarray}}

\newcommand{\ee}{\end{eqnarray}}

\newcommand\xleftrightarrow[1]{%
  \mathbin{\ooalign{$\,\xrightarrow{#1}$\cr$\xleftarrow{\hphantom{#1}}\,$}}
}

\usepackage[normalem]{ulem}

\usepackage[export]{adjustbox}

\providecommand{\sorthelp}[1]{}

\begin{document}

\preprint{APS/123-Q ED}

\title{The parity-odd Intrinsic Bispectrum}

\newcommand{\IPMU}{Kavli Institute for the Physics and Mathematics of the Universe (WPI), The University of Tokyo Institutes for Advanced Study (UTIAS), The University of Tokyo, Kashiwa, Chiba 277-8583, Japan}

\author{William R. Coulton}
\affiliation{\IPMU}

\date{\today}

\begin{abstract}
At linear order the only expected source of a curl-like, B mode, polarization pattern in the cosmic microwave background (CMB) is primordial gravitational waves. At second-order B modes are also produced from purely scalar, density, initial conditions. Unlike B modes from primordial gravitational waves, these B modes are expected to be non-Gaussian and not independent from the temperature and gradient-like polarization, E mode, CMB anisotropies. We find that the three point function between a second-order B mode and two first-order T/E modes is a powerful probe of second-order B modes and should be detectable by upcoming CMB experiments. We focus on the contribution to the three point function arising from non-linear evolution and scattering processes before the end of recombination as this provides new information on the universe at $z> 1000$. We find that this contribution can be separated from the other contributions and is measurable at $\sim 2.5 \sigma$ by CMB experiments with noise levels of $\sim 1 \mu$Karcmin and delensing efficiencies $\ge 90\%$, such as the proposed PICO satellite. We show that approximately half of the total signal arises from non-linearly induced vector and tensor metric perturbations, as evaluated in the Newtonian gauge. This bispectrum is a unique probe of these perturbations in the CMB, as their contribution to the power spectrum is suppressed. An important feature of this bispectrum is that the detectability will increase with decreasing experimental noise, in the absence of primordial B modes, provided that delensing efficiencies improve in parallel. 
\end{abstract}

\pacs{Valid PACS appear here}
\maketitle

\section{Introduction}
\label{sec:intro}
Precision measurements of the cosmic microwave background (CMB) have shown that the primordial universe was highly homogeneous and isotropic with small, Gaussian density (scalar) fluctuations \citep{Fixsen_1996,bennett2012,planck2016-l01}. The origin of the these fluctuations remains an open question in cosmology. Searching for vorticity (vector) and gravitational wave (tensor) fluctuations presents an interesting avenue for answering this question, as one promising theory for their origin, called inflation, predicts that the density fluctuations should be accompanied by a background of tensor perturbations\citep{Starobinskij1979,Rubakov1982,Abbott1984,Fabbri1983}. Primordial vectors, without sources, decay and are thus expected to be unobservable \citep{Hawking_1996}. Primordial gravitational waves imprint a distincive curl-like pattern of CMB polarisation anisotropies, B modes \citep{Kamionkowski_2016}. With the aim of detecting this signal, many existing and upcoming CMB experiments will make precision measurements of the CMB polarization anisotropies \citep{Ade2019,Abazajian2016,BICEP2018,Suzuki2018,Hanany2019}.

The scalar perturbations have been found to be very small, with deviations at the level of 1 part in $10^5$, and, whilst not yet detected, the vector and tensor perturbations are constrained to be at least equally small\citep{Fixsen_1996,bennett2012,planck2016-l01}. The smallness of the perturbations means their evolution is accurately described by linear perturbation theory, through which we can compute the CMB power spectra\citep{Bond_1987,Ma_1995,Seljak_1996}. At leading order in perturbation theory, the scalar, vector and tensor perturbations do not mix \cite{Kamionkowski_1997,Hu_1997} and the scalar fluctuations cannot generate B-mode anisotropies \citep{Seljak_1997}. However this is not the case at higher orders: non-linear evolution mixes types of perturbations and enables scalars to produce B modes. The most well known example is B modes produced by gravitational lensing that has recently been been measured \citep{Zaldarriaga_1998,Lewis_2006,Fa_ndez_2020,Sherwin_2017,Wu_2019,Planck_XVII_2013}.  In this work we explore the statistical properties of B modes generated by the second-order evolution of the density fluctuations.

We identify four primary sources of second-order B-mode anisotropies: (1) B modes arising from metric tensor and vectors induced by the non-linear evolution of the scalar modes \citep{Mollerach_2004,Bartolo_2006,Bartolo_2007,Pitrou_2009,Pitrou_2007}; (2) B modes induced by scattering processes at second-order; \citep{Beneke_2010,Beneke_2011}; (3) B modes induced as the CMB photons propagate through the inhomogeneous universe \citep{Zaldarriaga_1998,Lewis_2006,Hu_2000,Lewis_2017,Pratten_2016}; and (4) B modes arising due to the non-linear relationship between brightness and temperature \citep{Pitrou_2010,Pitrou_2014}. The induced B-modes' power spectrum has been carefully characterized \citep{Fidler_2014}.

The second-order B modes are not independent from the temperature (T) and the gradient-like CMB polarization pattern, E modes, as they are all sourced by the same scalar perturbations. However, the B-mode anisotropies have a different parity to the T and E modes, which means that the cross power spectrum of B modes with T or E modes vanishes. Higher-order correlation functions do not necessarily vanish and in fact, the non-linear evolution of the perturbations ensures that higher-order correlation functions will be produced, even from initially Gaussian perturbations. The temperature and E-mode bispectrum, the harmonic space three-point function, induced by the non-linear evolution of the perturbations has been studied extensively \citep{Pettinari_2013,Pettinari_2014,Pettinari_2016,Lewis_2012} and is known as the intrinsic CMB bispectrum. In this work we extend those calculations to compute the bispectrum between second-order B-mode anisotropies and first-order temperature and E-mode perturbations, i.e., the BTT, BTE and BEE bispectra. Due to the odd parity of B modes, this bispectrum will have odd parity and hence we refer to it as the parity-odd intrinsic bispectrum.

We show that the parity-odd intrinsic bispectrum is a powerful probe of second-order B modes. Through correlating the B mode with the well measured T and E modes, the parity-odd bispectrum provides a method to extract the second-order B modes from the noise and would potentially allow a measurement of these anisotropies before their power spectrum can be measured. We split the intrinsic bispectrum into three contributions: contributions from second-order sources at and before recombination, sources (1) and (2);  contributions induced by second-order sources after recombination,  source (3); and the quadratic sources, source (4) \footnote{Note that this split is not perfect as in fact the redshift term is mixed into both parts}. We find that the bispectra from these terms are differently shaped allowing them to measured independently. This is interesting as the bispectrum produced by processes at and before recombination explicitly provides new information on this era. 

Furthermore, this bispectrum gains significant contributions from non-scalar metric fluctuations. Mechanisms of generating non-scalar metric fluctuations, both in standard ($\Lambda$CDM) and non-standard cosmologies, is an area of active theoretical interest and these modes can be a novel probe of new pre-recombination era physics \citep{Baumann_2007,Luca_2020,ali_2020,Saga_2019,Ota_2020,Tomohiro_2020}.  In this paper we show that the parity-odd bispectrum offers a path to detecting these non-scalar modes. We find that proposed CMB experiments, such as PICO \citep{Hanany2019}, can marginally detect the bispectrum sourced by processes at and before recombination in $\Lambda$CDM . 

\section{The parity-odd intrinsic bispectrum} \label{sec:intrinsicBis}

\begin{figure}[t!]
 \centering
  \includegraphics[width=.45\textwidth]{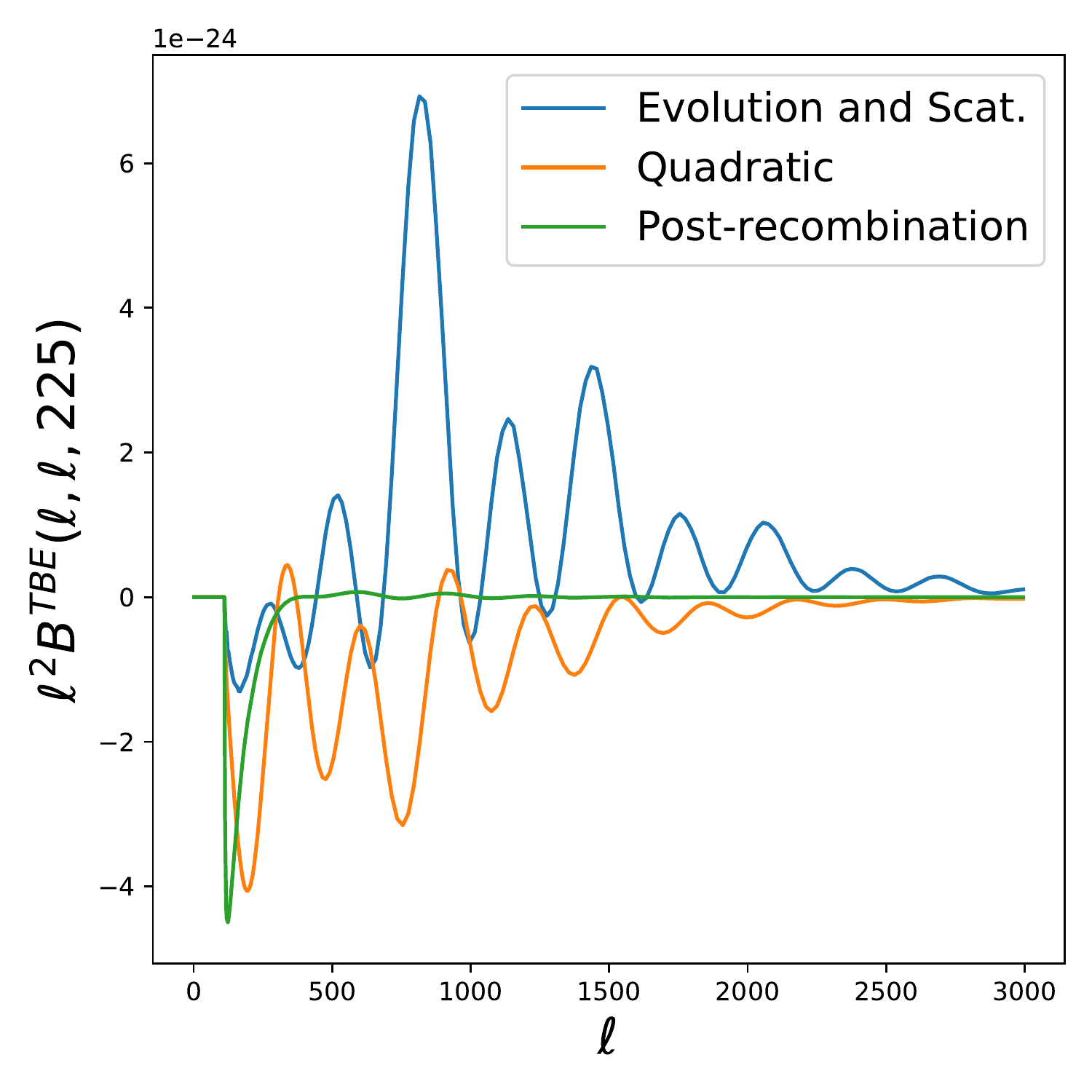}
  
\caption{A partially-squeezed-slice of the intrinsic bispectrum between a T, E and B field as a function of the scale of the T and B fields. This shows the different scale dependence of the three main components. Note that the geometric requirement that the bispectrum form closed triangles forces this slice of the bispectrum to be zero for $\ell\le112$. \label{fig:bispectrum_all}}
\end{figure}

\begin{figure}[t!]
  \centering
  \includegraphics[width=.45\textwidth]{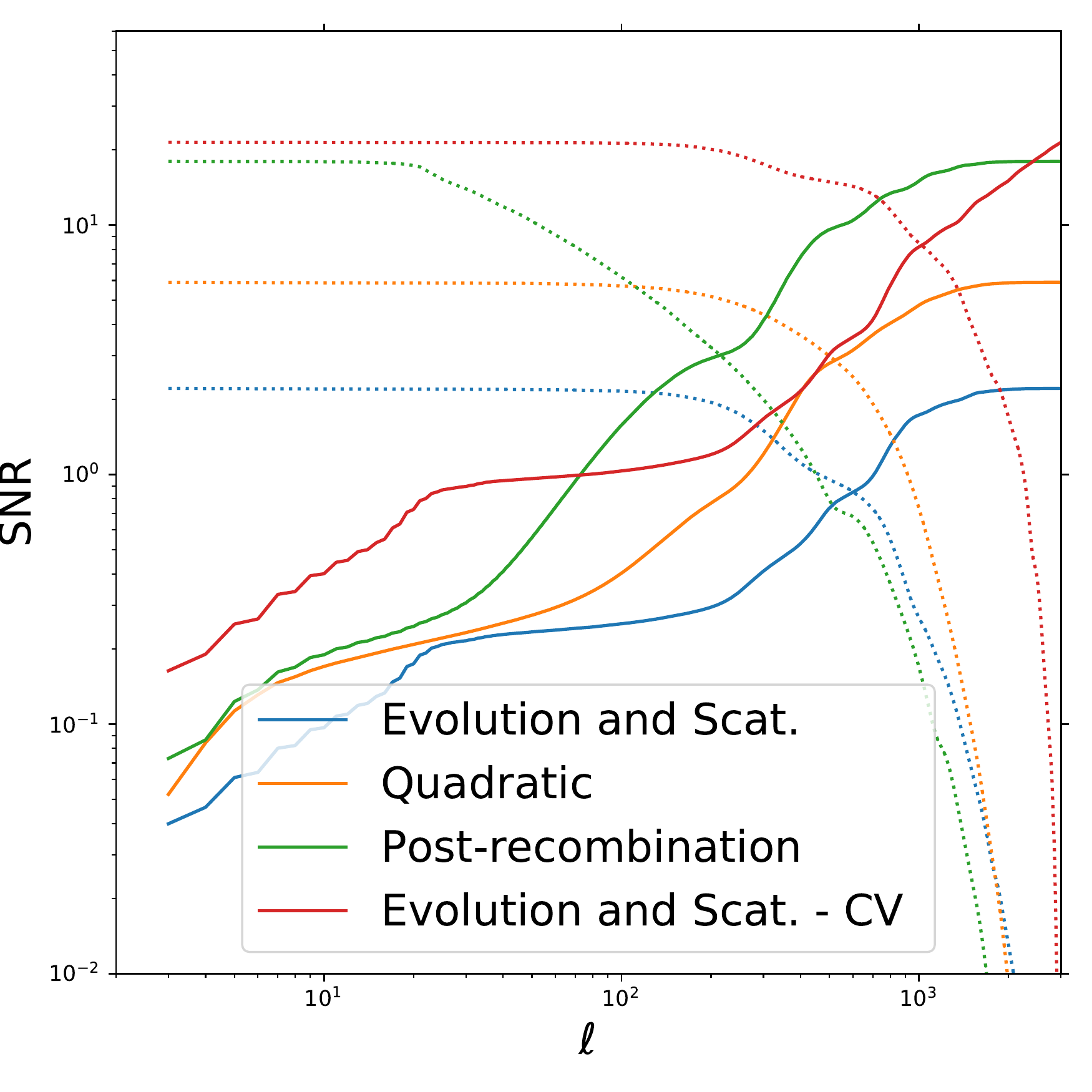} 
\caption{The signal to noise ratio (SNR) for the partiy odd bispectrum using a setup like PICO. This assumes f$_{\textrm{sky}}=0.8$ and that 90\% of the lensing power has been removed. In addition we plot the cosmic variance limited SNR, assuming 99\% delensing and r=0. In solid (dotted) lines we plot the SNR as a function of the maximum (minimum) scale included in the analysis. The SNR shown are obtained from joint constraint of all three bispectra. \label{fig:snr_parityOdd}}

\end{figure}
The derivation of the parity-odd bispectrum is analogous to the derivation of the parity-even bispectrum. As such, we present a brief overview and refer the reader to \citet{Pettinari_2013,Fidler_2014,Pettinari_2016} for details. 
The bispectrum is defined as $\langle a^X_{\ell_1 m_1} a^Z_{\ell_2 m_2} a^Z_{\ell_3 m_3} \rangle$, where $a^X_{\ell m}$ are the measured spherical harmonic coefficients of the T,E and B fields \citep{Spergel_1999}. A parity-odd bispectrum has $\ell_1+\ell_2+\ell_3 = $odd.
Analytically these coefficients are given by
\begin{align}
a^X_{\ell m}= \int \frac{\mathrm{d}^3 \mathbf{k}}{(2\pi)^3} (-i)^\ell \sqrt{\frac{4\pi}{2\ell+1}} \Theta_{X,\ell m}(\mathbf{k})e^{i \mathbf{k}\cdot \mathbf{x_0}},
\end{align} 
where $ \Theta_{X,\ell m}(\mathbf{k})$ is the temperature perturbation expressed in the Fourier and spherical harmonic basis and $X$ denotes the T,E and B components. Using the bolometric temperature definition \citep{Pitrou_2010,Pitrou_2014}, the temperature perturbation is related to brightness perturbation, $\Delta_{X}$, by
\begin{align} \label{eq:tempDef}
(\delta+\Delta)_X= (\delta+\Theta)^4_{X},
\end{align}
The brightness perturbation is a moment of the photon distribution function and is computed by solving the Boltzmann equations. It can be related to the initial primordial fluctuations, $\Phi(\eta_{\rm{in}},\mathbf{k})$, (assuming there are initially no vector or tensor modes) by
\begin{align}
& \Delta_{X,\ell m}(\eta_0,\mathbf{k}) =  \mathcal{T}^{(1)}_{X,\ell m}(\eta_0,\mathbf{k}) \Phi(\eta_{\rm{in}},\mathbf{k}) 
+\int\frac{ \mathrm{d}^3\mathbf{k'}}{(2\pi)^3}{ \mathrm{d}^3\mathbf{k''}} \nonumber \\ & \times \delta^{(3)}(\mathbf{k}+\mathbf{k'}+\mathbf{k''})\mathcal{T}^{(2)}_{X,\ell m}(\eta_0,\mathbf{k'},\mathbf{k''},\mathbf{k}) \Phi(\eta_{\rm{in}},\mathbf{k'}) \Phi(\eta_{\rm{in}},\mathbf{k''}),
\end{align} 
where $\mathcal{T}^{(i)}_X$ is the i$^{th}$ order transfer function for component X, $k$ is the comoving wavevector and $\eta$ the comoving time.

This quadratic coupling of Gaussian, primordial fluctuations leads to the following bispectrum
\begin{widetext}
\begin{align}\label{eq:bispectrumEvo}
&\langle a^X_{ \ell_1 m_1} a^Y_{\ell_2 m_2} a^B_{ \ell_3 m_3} \rangle_{\text{scat+evol}}=
 \sum \limits_{L_1,L_3,m}   \frac{ (-1)^{\ell_2}  }{4^3}\int \mathrm{d}r r^2 \prod\limits_{i=1}^3 \left[ \int \frac{\mathrm{d}k_i (-i)^{\ell_i}}{(2\pi)^3}  \sqrt{\frac{4\pi}{2\ell_i+1}} \right]  32\pi (2\pi)^3  (i)^{L_1+\ell_2+L_3}  \mathcal{I}_{m_1m_2m_3}^{\ell_1 \ell_2\ell_3}\mathcal{I}_{000}^{L_1 \ell_2L_3}\mathcal{I}_{000}^{\ell_1 L_1 |m|}
      \nonumber \\
    &\times \mathcal{I}_{m0 -m}^{\ell_3 L_3 |m|} \mathcal{J}^{\ell_1\ell_3\ell_2}_{L_3 L_1|m|} (2L_1+1)(2L_3+1) \mathcal{T}^{(1)}_{X \ell_1}({k_1})  \mathcal{T}^{(1)}_{Y \ell_2}({k_2})   P(k_1)P(k_2) 2 \mathcal{T}^{(2)}_{B \ell_3 m}({k_1},{k_2},{k_3}) j_{L_1}(rk_1)j_{L_2}(rk_2)j_{L_3}(rk_3) 
\end{align}
\end{widetext}
where \begin{align}
\mathcal{I}_{m_1m_2m_3}^{\ell_1 \ell_2\ell_3}=\begin{pmatrix}
    \ell_1& \ell_2 & \ell_3 \\ m_1 & m_2 &m_3
    \end{pmatrix} \text{, }  \mathcal{J}^{\ell_1\ell_3\ell_2}_{L_3 L_1|m|}  =\begin{Bmatrix}
    \ell_1 & \ell_3 & \ell_2 \\ L_3 & L_1 & |m|
    \end{Bmatrix}
\end{align}
are the Wigner 3j and 6j matrices,  $P(k)$ is the primordial power spectrum, and $j_\ell(x)$ are the spherical Bessel functions. The Bessel functions project the 3D, Fourier space fluctuations onto the surface of the sphere. The Wigner symbols contain the geometrical couplings of the different Fourier modes. 
The transfer functions, obtained by numerically solving the Boltzmann equations  using the SONG and CLASS codes\citep{lesgourgues2011cosmic,Blas_2011,Pettinari_2013},  encode how the initial perturbations evolve and couple to source the CMB anisotropies.
We stop the numerical evolution after recombination, and thus this bispectrum primarily captures the scattering and non-linear evolution terms, sources (1) and (2), hence we call this the scattering and evolution contribution. Physically these sources arise as the quadratic coupling of two non-aligned scalar modes generates non-scalar modes. In the first case the non-linear coupling from the gravitational evolution generates non-scalar perturbations that then source B modes in an identical manner to linear non-scalar modes.  In the second case, scattering B modes arise due to effects including the doppler boosting by the electrons' bulk velocity and the modulation of the scattering rate by long wavelength density modes .

To solve the Boltzmann equation, we solve the transfer functions for the transformed variable
\begin{align} \label{eq:redshift}
\tilde{\Delta}_{X} = \left[\ln(1+\Delta) \right]_{X}.
\end{align}
This transformation simplifies the Boltzmann equations by refactoring the redshift term \citep{Huang_2013}. The transfer functions in Eq. \eqref{eq:bispectrumEvo} are in terms of this variable. To obtain the physical bispectrum we must transform back to $\Delta_{X}$. This transformation and the non-linear relation to the temperature perturbation in Eq. \eqref{eq:tempDef} lead to the second bispectrum contribution, hereafter the quadratic term,
\begin{widetext}
\begin{align}\label{eq:bispectrum}
&\langle a^X_{ \ell_1 m_1} a^Y_{\ell_2 m_2} a^B_{ \ell_3 m_3} \rangle_{\text{quad}} =
     -i\sqrt{ \frac{ (2\ell_1+1)(2\ell_2+1)(2\ell_3+1)}{4\pi}} 
       \mathcal{I}_{m_1m_2m_3}^{\ell_1 \ell_2\ell_3}     
     \left(\frac{1 - (-1)^{\ell_3+\ell_1+\ell_2}}{2}\right)\Big[ \mathcal{I}_{02-2}^{\ell_1 \ell_2 \ell_3}
C^{XT}_{\ell_1}C^{YE}_{\ell_2} -\mathcal{I}_{02-2}^{\ell_2 \ell_1 \ell_3} C^{XE}_{\ell_1}C^{YT}_{\ell_2}  \Big] ,
\end{align}
\end{widetext}
where $C^{XY}_{\ell}$ is the power spectrum.

The final contribution arises from post-recombination sources and is comprised of the lensing, emission-angle and time-delay effects. It is more efficient to evaluate these sources separately, rather than evolve the Boltzmann equations \citep{Fidler_2014}, hence we compute this contribution using the results from \citet{Goldberg_1999,Hu_2000,Lewis_2017,Pratten_2016}. These contributions, hereafter post-recombination terms, are given by 
\begin{widetext}
\begin{align}\label{eq:bispectrum}
&\langle a^X_{ \ell_1 m_1} a^Y_{\ell_2 m_2} a^B_{ \ell_3 m_3} \rangle_{\text{post-recomb.}} = 
   -i  \sqrt{ \frac{(2\ell_1+1)(2\ell_2+1)(2\ell_3+1)}{4\pi} }\left\{  \mathcal{I}_{m_1m_2m_3}^{\ell_1 \ell_2\ell_3}      \mathcal{I}_{0-22}^{\ell_1 \ell_2\ell_3}  \left(\frac{1 - (-1)^{\ell_3+\ell_1+\ell_2}}{2}\right) \right. \nonumber \\ 
&  \left. \times\left[ \sqrt{\frac{(\ell_2+2)!}{(\ell_2-2)!}} C^{\psi_d Y}_{\ell_2}C^{\psi_{\zeta}X}_{\ell_1}+C^{\psi_{d} X}_{\ell_1}C^{EY}_{\ell_2}-\frac{1}{2}\left[\ell_1(\ell_1+1)+\ell_2(\ell_2+1)-\ell_3(\ell_3+1) \right] C^{\phi X}_{\ell_1}C^{EY}_{\ell_2} \right] + \ell_1,X \xleftrightarrow{} \ell_2,Y  \right\},
\end{align}
\end{widetext}
where  $\phi$,$\psi_d$ and $\psi_\zeta$ are the lensing, emission and the polarization potentials \citep{Lewis_2017}. The first two terms are the emission-angle and time-delay effects and the final term is the lensing contribution. We neglect post-Born effects as they are higher-order contributions \citep{Pratten_2016}.  

As shown in Figure \ref{fig:bispectrum_all} the different contributions have distinct scale dependence. For this slice, all three bispectra are important, whilst for most slices one of the terms dominates, e.g., squeezed configurations, $\ell_1\sim\ell_2 \gg\ell_3$, with large-scale temperature modes are dominated by the post-recombination terms.

\section{Detectability}
New information encoded at recombination is primarily contained in the evolution and scattering term, as such we consider measuring the three components of the intrinsic bispectrum separately. We forecast the bispectrum's detectability for a CMB experiment with a $6.2$ arcmin FWHM beam and Q/U map noise levels of $0.9\mu$Karcmin, consistent with a PICO-like experiment  \citep{Hanany2019}. We assume delensing can remove $90\%$ of the lensing power, which should not be beyond the reach of these experiments \citep{Diego_Palazuelos_2020}. We delens the lensing-induced bispectrum assuming that the delensing procedure purely reduces the amplitude of the lensing potential \footnote{This is trivially implemented by rescaling the terms that depend on the lensing potential by the map level delensing factor. e.g by rescaliing $C^{\phi T}$ by the square root of the delensing power efficiency}.

In Figure \ref{fig:snr_parityOdd} we plot the signal to noise ratio (SNR). We see that the quadratic and post-recombination bispectra can be strongly detected, whilst the evolution and scattering bispectrum is detectable at the $\sim 2.5 \sigma$ level. The strongest bispectrum is that from the post-recombination sources and is dominated by the lensing-ISW contributions \citep{Lewis_2011}.

An interesting aspect of this signal is that the SNR of all bispectra will continue to improve until providing the lensing power can be removed until a noise floor is reached in the B-mode power spectrum.  Currently it is unclear what will set the fundamental B-mode floor and thus the fundamental limit on the bispectrum SNR. It is likely that this floor will arise from either emission-angle effects \citep{Lewis_2017}, post-Born lensing effects \citep{Pratten_2016}, B-modes from reionization \citep{Dvorkin_2009,Zahn_2005,Gnedin_2001,Roy_2021} or limitations in delensing techniques \citep{lizancos_2020} which could limit the detectability of the intrinsic bispectrum to SNR$\sim20$-$100$. Assuming this limits delensing to $99\%$ we plot the cosmic variance limited SNR in Figure \ref{fig:snr_parityOdd}. Our calculations neglect the non-Gaussian contributions to the variance which may be important for high SNR measurements.

The main signal contaminant that would need to be mitigated would be the bispectrum from Galactic foregrounds \citep{Coulton_2019}. Contamination from the Galaxy can be removed by utilizing the different foreground spectral properties \citep{Planck_XI_2018} and, as removing these foregrounds is a major challenge for primordial B-mode searches, experiments are designed with sufficient frequency coverage to enable this separation \citep{s4collaboration2020cmbs4}. Note extra-galactic sources, whilst producing detectable temperature bispectra \citep{Crawford_2014,Coulton_2018}, are not expected to produce significant polarized, parity-odd bispectra \citep{lizancos2021delensing}. 

One concern is the impact of delensing on the bispectrum signal itself. \citet{Coulton_2020} showed that delensing does not introduce a bispectrum, however delensing can bias non-zero measurements of primordial scalar bispectra towards zero. \citet{Coulton_2020} also showed that such biases can be removed by the careful construction of the lensing potential, used in the delensing process. We defer the investigation of whether delensing introduces any biases and of whether such bias can be removed with similar techniques to future work as this requires addressing the challenging problem of simulating the intrinsic bispectrum.
\begin{figure}[t]
    \centering
  \includegraphics[width=.45\textwidth]{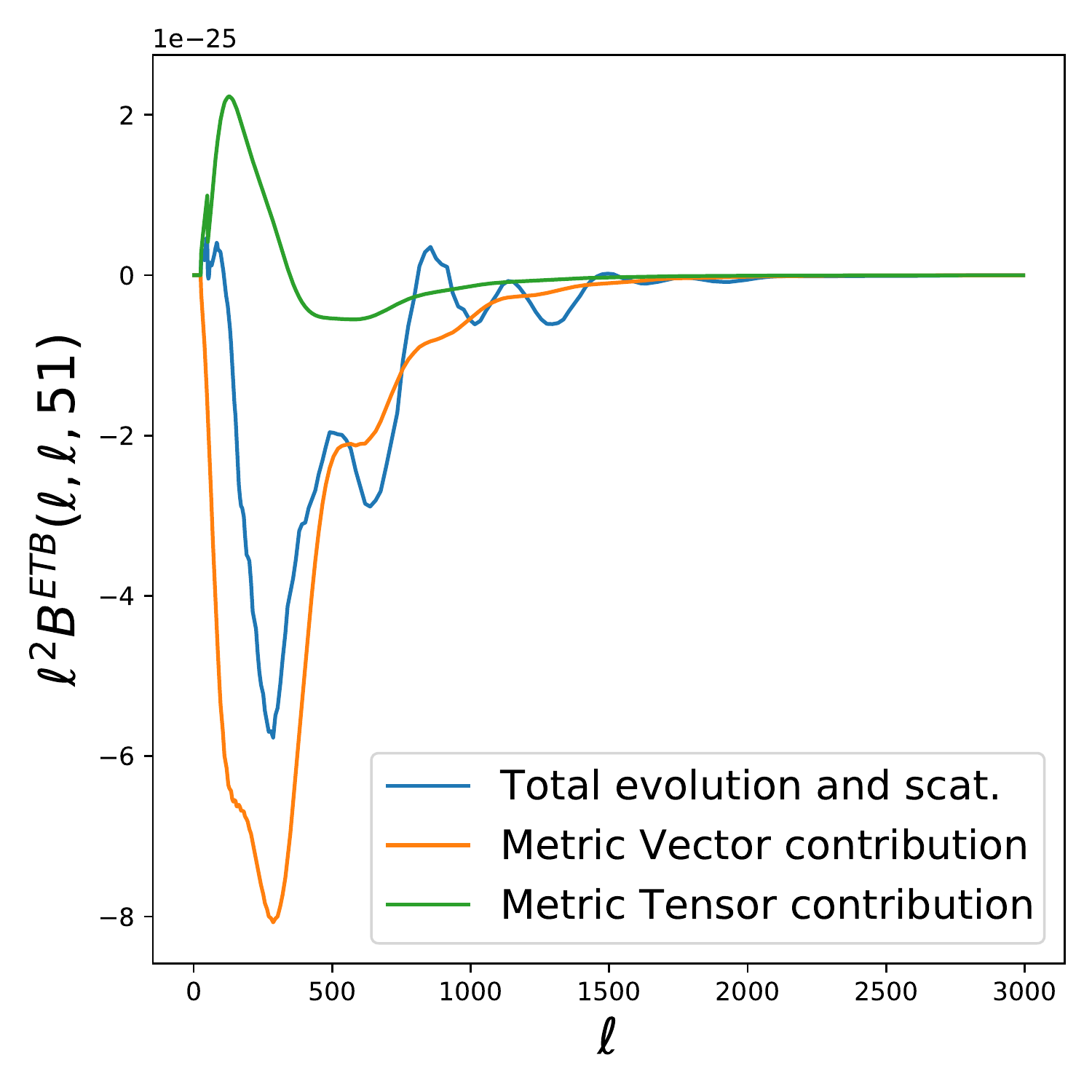}
 
\caption{A partially-squeezed-slice of the intrinsic bispectrum arising from the Newtonian-gauge metric vector and tensor modes. We plot the total evolution and scattering bispectrum for comparison. \label{fig:bispectrum_V_T}}
\end{figure}

\begin{figure}[t]
  \centering
  \includegraphics[width=.45\textwidth]{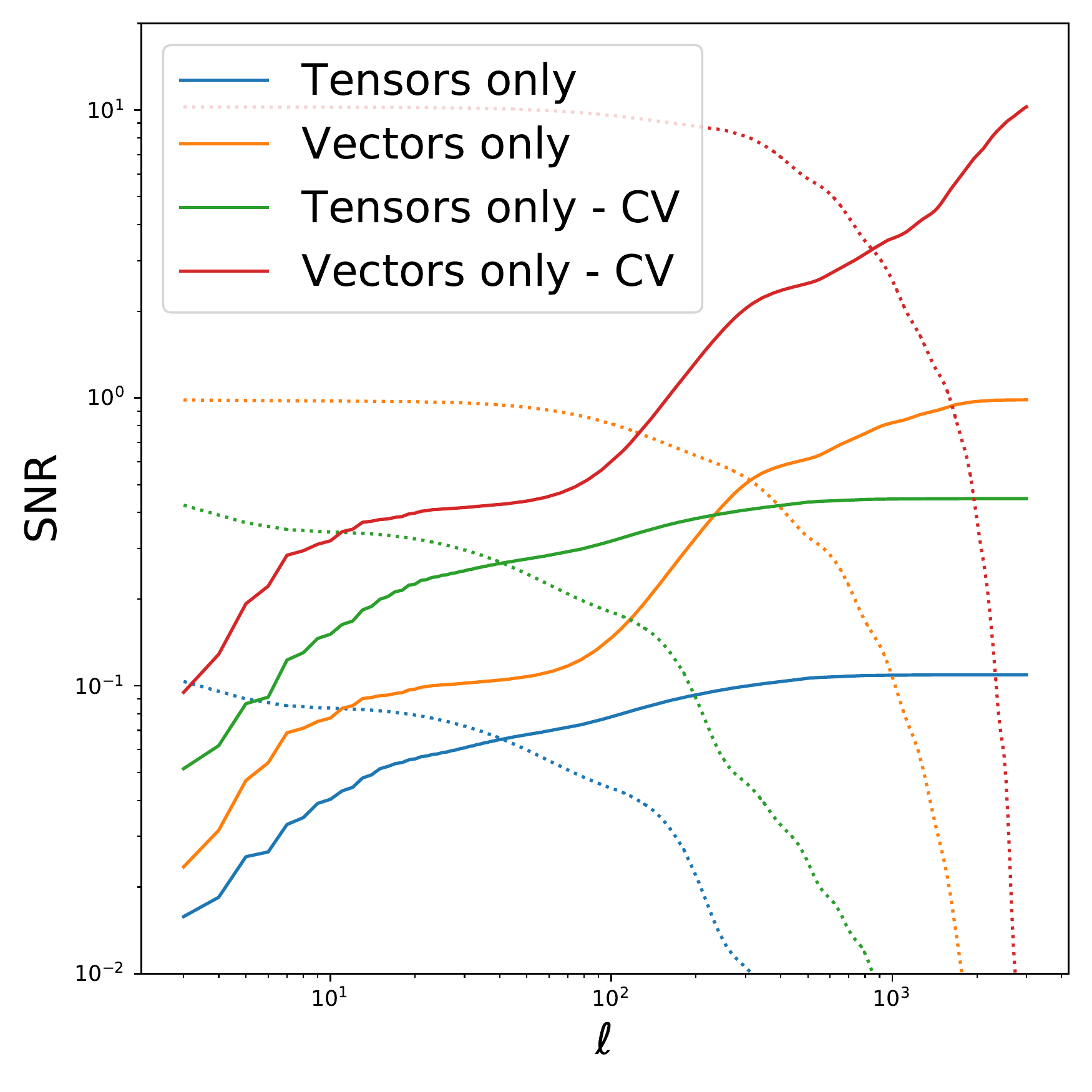} 
\caption{The SNR for the Newtonian-gauge metric vector and tensor contributions to the parity-odd bispectrum using a setup like PICO and for the CV case, assuming 99\% delensing. In solid (dotted) lines we plot the SNR as a function of the maximum (minimum) scale included in the analysis. We marginalise over the quadratic and post-recombination bispectra.  \label{fig:snr_parityOdd_VT}}

\end{figure}
 
\section{Probe of non-scalar metric modes}
Second-order B modes are only sourced by non-scalar modes and so are a powerful probe of non-scalar perturbations.  A particularly interesting source of second-order B modes is the second-order metric  perturbations as a measurement of these terms would probe vorticity and gravitational wave modes on cosmological scales \citep{Mollerach_2004}. At second-order, the metric perturbations are gauge dependent. Following \citet{Pitrou_2007,Pitrou_2009}, we compute the B modes sourced by a combination of the perturbations that form gauge invariant vector and tensor perturbations. In the Newtonian gauge, in which we work, these gauge invariant perturbations reduce to the metric vector and tensor perturbations. 
Despite this complication, \citet{Ali_2021} showed that the induced second-order gravitational waves, in matter domination, are the same in a wide range of gauges and the B modes sourced by our gauge-invariant tensor perturbation correspond to the B-modes from induced gravitational waves. 

We compare the amplitudes of these contributions to the total evolution and scattering contribution in Figure \ref{fig:bispectrum_V_T}.   These bispectra have non-trivial amplitudes and so in Figure \ref{fig:snr_parityOdd_VT} we explore whether they could be measured. We find that the vector bispectrum is significant and potentially also detectable in the near future. Unfortunately the bispectrum from tensor modes is significantly smaller and beyond upcoming experiments.  
\section{Utility for parameter inference}

By inspecting Eq. \eqref{eq:bispectrum}, we can immediately see that the intrinsic bispectrum could break degeneracies between CMB parameters. Most obviously we see that the parity-odd amplitude depends on $A_s^2 e^{-3\tau}$ compared to the CMB power spectrum, which on small scales, has a $A_s e^{-2\tau}$ dependence. Unfortunately, currently proposed experiments will not be able to improve significantly their constraining power on parameters such as $\tau$ and $A_s$, as the parity-odd bispectrum will be measured at a low significance compared to the power spectrum. However,  as the parity-odd bispectrum SNR increases with delensing capability, in the longer term it could be used to break such power spectrum degeneracies. 

\section{Bias for primordial tensor-scalar non-Gaussianity searches}
As seen in Figure \ref{fig:snr_parityOdd}, the intrinsic parity-odd bispectrum gains significant contributions from B modes on $\ell$ scales of a few hundred. This is at different to scales compared to those responsible for primordial tensor-scalar non-Gaussianities as primordial tensors rapidly decay within the horizon \citep{Meerburg_2016,Duivenvoorden_2020}. This means that these bispectra are likely highly orthogonal. For upcoming surveys it is therefore not expected that the intrinsic bispectrum will be a significant bias to searches for primordial tensor-scalar non-Gaussianities. A caveat of this conclusion is that this work has only included contributions to the intrinsic bispectrum from recombination. It is expected that there will contributions to this bispectrum from reionization that will peak on larger scales \citep{Mollerach_2004}. These contributions have the potential to bias primordial tensor-scalar measurements. We defer a thorough analysis of this term to future work.

\section{Conclusions}
In this work we calculated, for the first time, the parity-odd intrinsic bispectrum. This bispectrum, between a second-order B mode  and two first-order T/E modes, will be detectable with CMB missions like the proposed PICO mission. We separated the bispectrum into three physical contributions: late times sources, quadratic sources, and evolution and scattering sources. The later source provides novel information about $z> 1000$ physics. All three contributions should be just measurable with a PICO-like mission. By correlating the weak, second-order B modes with two first-order T/E modes we can pull this signal from the noise. This process means that the bispectrum is a more powerful probe than the power spectrum of second-order B modes \citep{Fidler_2014}.

The parity-odd bispectrum gains significant contributions from second-order, Newtonian-gauge metric-vector perturbations and would be a unique probe of these modes. Second-order, Newtonian-gauge metric tensor modes also contribute, though detecting these will be beyond the capabilities of upcoming CMB experiments. In principle measurements of this bispectrum could break parameter degeneracies in the CMB power-spectrum, however in practice this will be difficult due to the significantly lower SNR of the bispectrum. It is not expected that the intrinsic bispectrum will bias PNG searches, due to the different scale dependence.

\begin{acknowledgments}
\textit{Acknowledgements} The author is incredibly grateful to Guido W. Pettinari,  Christian Fidler and the other contributors and maintainers of SONG. Their efforts, in particular the excellent documentation, hugely facilitated this project. Further, I would like to thank Anthony Challinor, Atsuhisa Ota, Antony Lewis, Daan Meerburg and Christian Fidler for highly productive discussions.  W.R.C. acknowledges support from the UK Science and Technology Facilities Council (grant number ST/N000927/1), the World Premier International Research Center Initiative (WPI), MEXT, Japan and the Center for Computational Astrophysics of the Flatiron Institute, New York City. The Flatiron Institute is supported by the Simons Foundation. 
\end{acknowledgments}
\bibliographystyle{apsrev.bst}
\bibliography{intrinsic,planck_bib}

\end{document}